%Paper: hep-ph/9504273
%From: Stephan Titard <stephan@nantes.ft.uam.es>
%Date: Mon, 10 Apr 95 19:52:18 GMT+0100

%%%%%%%%%%%%%%%%%%%%%%%%%%%%%%%%%%%%%%%%%%%%%%%%%%%%%%%%%%%%%%%%%%%%%%%%%
% sty.tex
% version 10/4/95
% sent by S.Titard
%%%%%%%%%%%%%%%%%%%%%%%%%%%%%%%%%%%%%%%%%%%%%%%%%%%%%%%%%%%%%%%%%%%%%%%%%
\magnification 1200
\input epsf
\def\pmbf#1{\setbox0=\hbox{#1}%
\kern-.025em\copy0\kern-\wd0
\kern.05em\copy0\kern-\wd0
\kern-.025em\copy0\kern-\wd0
\kern-.04em\copy0\kern-\wd0
\kern.08em\box0 }
\rightline {FTUAM $\oldstyle  10$/$\oldstyle 95$, April $\oldstyle 1995$}
\vskip.5cm
\centerline {\bf HEAVY QUARKONIUM SYSTEMS}
\centerline {\bf AND}
\centerline {\bf NONPERTURBATIVE FIELD CORRELATORS}
\vskip 1.5cm
\centerline {{\bf Yu. A. Simonov}\footnote*{e-mail:simonov@vxitep.itep.ru}}
\vskip .2cm
\centerline{\sl ITEP,}
\centerline {\sl B. Cheremushkinskaya {$\oldstyle 25$},}
\centerline {{$\oldstyle 117259$}, Moscow, Russia}
\vskip .4cm
\centerline {{\bf S. Titard}\footnote{**}{\rm e-mail: stephan@nantes.ft.uam.es}
 {\sl and}
{\bf
 F. J. Yndur\'ain}\footnote{***}{\rm e-mail:fjy@delta.ft.uam.es}}
\vskip .2cm
\centerline{\sl Departamento de F\'{\i}sica Te\'orica, C-XI,}
\centerline{\sl Universidad Aut\'onoma de Madrid, Canto Blanco,}
\centerline{E-$\oldstyle 28049$, Madrid, Spain.}
\vskip 1.5cm
{\bf Abstract}.

Bound states of heavy quarks are considered. Using the path integral formalism
we are able to rederive, in a gauge invariant way,
the Leutwyler-Voloshin short distance analysis
 as well as a long distance linear potential. At all distances we describe the
states in terms of nonperturbative field correlators, and we include radiative
corrections
at short and intermediate distances. For intermediate distance
 states (particularly $b\bar {b}$ with $n=2$) our results improve,
qualitatively and
 quantitatively, standard analyses, thanks mostly to being
able to take into account the finiteness of the correlation time.
\vskip 2cm
Typeset with Plain \TeX

\vfill
\eject
Some time ago, Leutwyler$^{[1]}$ and Voloshin$^{[2]}$ showed that the gluon
condensate
 $\langle \alpha_s\,G^2 \rangle$ controls the leading nonperturbative effects
for heavy
 $q\bar {q}$ states {\it at short distances}. Their analysis was completed in
refs. 3,4,
in particular by extending it to spin dependent splittings and by including
relativistic
 and one loop radiative corrections, an essential ingredient in the analysis.
With
 these additions it was then shown that a consistent description
of states $n=1\,\,b\bar {b}$ and, to a lesser extent, $b\bar {b}$ states with
$n=2$
and $c\bar {c}$ states with $n=1$ could be obtained.\footnote*{$n$ is the
principal quantum number.
We will use standard atomic spectroscopic notation}

As already pointed out in refs.~1,2 the approach fails for large $n$. The
reason is that nonperturbative
contributions grow like $n^6$, quickly getting out of hand. For example, for
the spin-independent spectrum
 we have,
$$M(n,l)=2m \Bigl\{ 1-{{C_F{\widetilde {\alpha}_s}^2}\over {8n^2}}$$
$$-\left[ \log{{n\mu}\over {mC_F{\widetilde {\alpha}_s}}}+\psi (n+l+1)\right]
\,{{C_F^2\beta_0 {\widetilde {\alpha}_s} \alpha_s^2}\over {8\pi n^2}}
+{{\pi \epsilon_{nl} n^6\,\langle \alpha_s\,G^2 \rangle }
\over {2(mC_F{\widetilde {\alpha}_s})^4}}\Bigr\}.
 \eqno (1) $$
Here $l$ is the angular momentum, $m$ is the pole mass of the quark,
 $\beta_0=(33-2n_f)/3$, $C_F={{4}\over {3}}$ and ${\widetilde {\alpha}}_s$
 embodies part of the radiative corrections:
$$\widetilde {\alpha}_s(\mu )=
\Bigl\{1+\left[{{93-10n_f}\over {36}}+{{\gamma_E \beta_0}\over
{2}}\right]{{\alpha_s}\over {\pi }}
\Bigr\}.$$
Here $\mu$ is the renormalization point. Finally, in the leading
nonperturbative approximation,
{\it and assuming a constant gluon condensate density},
$$\langle G_{\mu \nu} (x)G_{\mu \nu}(y)\rangle \simeq
\langle G_{\mu \nu} (0)G_{\mu \nu}(0)\rangle \equiv \langle G^2 \rangle,$$
\noindent the $\epsilon_{nl}$ are numbers of order unity:
\vskip .2cm

$\epsilon_{10}={{624}\over {425}},\,\epsilon_{20}={{1\,051}\over {663}},\,
\epsilon_{30}={{769\,456}\over {463\,239}},\,\epsilon_{21}={{9\,929}\over
{9\,944}},\dots$
\vskip 0cm
 \rightline {(2)}
\vskip .2cm

\noindent We will not consider in this note relativistic corrections.

Clearly, the nonperturbative correction in Eq.(1) blows up very quickly for
increasing $n$
 making the method totally unsuitable already for $b\bar {b}$ with $n=3$, and
$c\bar {c}$
 with $n=2$. The way out of this difficulty found in most of the literature is
to use
a phenomenological potential,
$$\sigma r,\,\sigma^{{1}\over {2}}\,\simeq 0.45\,{\rm MeV}. \eqno (3) $$
Now, and although a linear potential yields a correct description of {\it long}
distance $q\bar {q}$
forces, the methods lack rigour in that, as is well known$^{[6;1,2]}$, a linear
potential
 is incompatible with known QCD results {\it at short distances}, where indeed
it does not
represent a good approximation of {\it e.g.}, the nonperturbative part of (1).

Because of this it is desirable to develop a framework which, in suitable
limits, implies
both the Leutwyler-Voloshin short distance results as well as the long distance
description
in terms of a linear potential. This framework is an elaboration of that
developed in
 refs.~7, where it was shown from first principles
how one can derive a linear potential from first principles.
In the present note we explore the short and
intermediate distances, where we rederive the Leutwyler-Voloshin
 description, {\it improved} both by getting better agreement with experiment
for the states where it is
 valid, and extending its range of applicability to intermediate distances. The
reason for this improvement lies
in that our treatment includes the {\it nonlocal} character of the gluonic
condensate
in the form of a finite correlation time, $T_g$. The results of refs. 1 to 4
are then recovered
 in the limit $T_g\rightarrow \infty$.

The nonlocal condensates have been considered previously$^{[9,10]}$, in
particular with the aim of
finding this correlation time. In this note we improve upon the treatment of
refs. 9,10 first by
 using a Lorentz and gauge invariant path integral formulation which would
allow us, if so wished,
to incorporate relativistic and spin effects\footnote{**}
{The methods to acomplish this would be like the ones developed in the fifth
paper of ref. 7 and in ref. 8}.
 Secondly, we incorporate radiative corrections which are essential to give a
meaning
to parameters like $\alpha_s (\mu )$ or $m$.
\vskip 1.5cm
{\bf Description of the method.}
\vskip .2cm
The method uses the path integral formalism. Because in this note we are only
interested in
nonrelativistic, spin independent splittings, we start directly with the
 nonrelativistic $q\bar {q}$
Green's function$^{[7,8]}$. For large time $T$,
$$G(x,\bar {x};y,\bar {y})=4m^2\,{\rm e}^{-2mT}
\int {\cal D}{\bf z}\,{\cal D}\bar {\bf z}\,{\rm e}^{-(K_0+\bar
{K}_0)}\,\langle W(C)\rangle,
\eqno (4) $$
$$K_0={{m}\over {2}}\int^T_0{\rm d}t\,\dot {\bf z}(t)^2,\,
\bar {K_0}={{m}\over {2}}\int^T_0{\rm d}t\,\dot{\bar{\bf z}} (t)^2.$$
\noindent $W(C)$ is the Wilson loop operator corresponding to the closed
 contour $C$ which includes the $q$, $\bar {q}$ paths. $W(C)$
should also include initial and final parallel transporters, $\Phi (x,\bar
{x}),
\Phi (y,\bar {y})$ with {\it e.g.},
$$\Phi (x,\bar {x})={\rm P}\,\exp {\rm i}g\int^x_{\bar {x}}{\rm
d}z_{\mu}\,B_{\mu}(z), \eqno (5) $$
\noindent $P$ denoting path ordering. Actually, we can omit the parallel
transporters,
and at the same time avoid problems with the renormalization of the Wilson loop
by choosing $x=\bar {x},\,y=\bar {y}$, which will prove sufficient for our
purposes.

In order to take into account the nonperturbative character of the interaction
we split the gluonic
field $B$ as
$$B_{\mu}=b_{\mu}+a_{\mu}.$$
\noindent The separation will be such that, by definition,
the vacuum expectation value of Wick ordered products of $a_{\mu}$
 vanishes, so the correlator may be written in terms solely of $b_{\mu}$:
$$\langle G (x)G(y)\rangle \rightarrow
\langle G_b (x)G_b(y)\rangle, \eqno (6) $$
and $G_b$ is constructed with only the $b_{\mu}$ piece of $B_{\mu}$.
 One may expand in powers of the $b_{\mu}$ and
thus write the Wilson loop average as
$$\langle W(C)\rangle=\int {\cal D}a\,{\rm P}\exp {\rm i}g\int_C {\rm
d}z_{\mu}\,a_{\mu}$$
$$+\left({{{\rm i}g}\over {2!}}\right)^2 \int {\cal D}a\,
\int_C{\rm d}z_{\mu} \int_C{\rm d}z'_{\nu}\,
{\rm P}\Phi_a (z,z')b_{\mu}(z'){\rm P}\Phi_a (z',z)b_{\nu}(z')+\dots $$
$$\equiv W_0+W_2+\dots, \eqno (7)$$
where the transporter $\Phi_a$ is given by an equation like (5), but in terms
of
only the $a_{\mu}$ piece of $B_{\mu}$.

Let us discuss the first term in the r.h.s. of Eq. (7). Using the cluster
expansion we get
$$W_0=Z\exp(\phi_2+\phi_4+\dots),$$
$$\phi_2=-{{C_Fg^2}\over {8\pi^2}}\,\int \int{{{\rm d}z_{\mu}{\rm d}
z'_{\mu}}\over {(z-z')^2}}
, \eqno (8)$$
\noindent and regularization (to be absorbed in $Z$) is implied in this
integral.
 It should be noted that $\phi_2$ contains all ladder-type exchanges, and in
addition also
 "Abelian crossed" diagrams -those where the times of the vertices can be not
ordered,
 but where the color generators $t^c_{ik}$ are always kept in the same order.
 Because of this, all crossed diagrams (with the exception of the "Abelian
crossed")
are contained in $\phi_4$. It is remarkable that each term $\phi_{2n}$ in
 Eq.(8) sums up an infinite series of diagrams. In particular,
 and as we will see below, $\exp\phi_2$ contains all powers
 of $\alpha_s /v$ ($v$ being the velocity of the quarks) so the calculation
is exact in the nonrelativistic limit.

For heavy (and slow) quarks, (8) becomes,
$$\phi_2={{C_Fg^2}\over {4\pi^2}} \int^T_0{\rm d}t \int^T_0{\rm d}t'\,
{{1+\dot{\bf z}\,\dot{\bf z}'}\over {{\bf
r}^2+(t-t')^2}}=C_F\alpha_s\,r^{-1}\int^T_0{\rm d}t
+O(v^2), \eqno (9)$$
\noindent {\it i.e.}, a singlet one-gluon exchange potential, as expected.

We then turn to $W_2$. When expanding it in powers of $a_{\mu}$ one gets
typical terms like
$${\rm Tr}(t^{c_k}\,t^{c_{k-1}}\dots t^{c_1} t^{c_2} t^{c_2}\dots t^{c_k}
t^{a_1}\dots t^{a_n}\, t^{a}\,b^a_{\mu}\,t^{a_n} \dots t^{a_1}\,b^c_{\nu}\,
t^c)$$
$$\rightarrow C_F^k\, {\rm Tr}(t^{a_1}\dots t^{a_n}\,
t^{a}\,b^a_{\mu}\,t^{a_n} \dots t^{a_1}\,b^c_{\nu}\, t^c\dots).$$
\noindent Because of the equality
$$t^c\,t^a\,t^c=-t^a/2N_C,$$
one obtains, for all exchanges in the time interval between the times of
$b_{\mu}(z)$ and
 $b_{\nu}(z')$ a factor $-1/N_C$ instead of the factor
$C_F$ that is found for other exchanges. As a result we may write
$W_2$ as
$$W_2={{({\rm i}g)^2}\over {2}}\int_C {\rm d} z_{\mu} \int_C {\rm d}z'_{\nu} \,
 \langle b_{\mu} b_{\nu} \rangle $$
$$\times \exp \left [ \int^T_{t_2} {\rm d}t\,V^{(S)}_C+\int^{t_1}_0 {\rm
d}t\,V^{(S)}_C+
\int^{t_2}_{t_1}{\rm d}t\,V^{(8)}_C \right] , \eqno (10) $$
with $V^{(S)}_C=C_F\alpha_s\,/r,\,V^{(8)}_C=-(1/N_C)\, \alpha_s\,/r$ the
singlet, octet potentials
 respectively. The relevance of the octet potential was already noted in refs.
1,2. It appears in
our derivation in a fully gauge invariant way, in connection with a gauge
invariant Green's function
 with the gauge invariant quantity $W_2$ as the kernel.
\vskip 1.5cm
{\bf Evaluation of} $W_2$
\vskip .2cm
One first uses the Fock-Schwinger gauge
 (see ref. 6 for details, including a modification of this gauge) to write,
$$\int_C {\rm d} z_{\mu} \int_C {\rm d}z'_{\nu} \, \langle b_{\mu} b_{\nu}
\rangle=
\int {\rm d}\sigma_{\mu \rho}{\rm d} \sigma'_{\nu \lambda}
\langle G_{b\mu \rho}(z)G_{b\nu \lambda}(z')\rangle, $$
and the ${\rm d}\sigma$ are surface differentials. Including also parallel
transporters,
 equal to unity in the F-S gauge, we find the gauge-invariant
 expression,
$$\int_C {\rm d} z_{\mu} \int_C {\rm d}z'_{\nu} \,\langle b_{\mu}b_{\nu}
\rangle =\int {\rm d}\sigma_{\mu \rho}{\rm d} \sigma'_{\nu \lambda}
\Bigl\{ \Phi(x_0,w)G_{b\mu \rho}\Phi(w,x_0)$$
$$\times  \Phi(x_0,w')G_{b\nu \lambda}\Phi(w',x_0) \Bigr\}.\eqno (11)$$
\noindent As was shown in ref. (7) $x_0$ may be chosen between $w$ and $w'$,
 up to additional contributions of order $b_{\mu }^4$, that we are neglecting
here.
 Then we divide the total time interval $T$ into three parts ({\it cf.} Fig.1):
$$(I)\,\,\,\,0\leq t \leq w'_4$$
$$(II)\,\,\,w'_4\leq t \leq w_4$$
$$(III)\,\,w_4\leq t \leq T.$$
\noindent Separating out the trivial {\it c.m.} motion we get, in regions ({\it
I}), ({\it III}),
the singlet Coulomb Green's function,
\vfill
\eject
\vskip .2cm
\epsfxsize=145truemm
\centerline{\epsffile{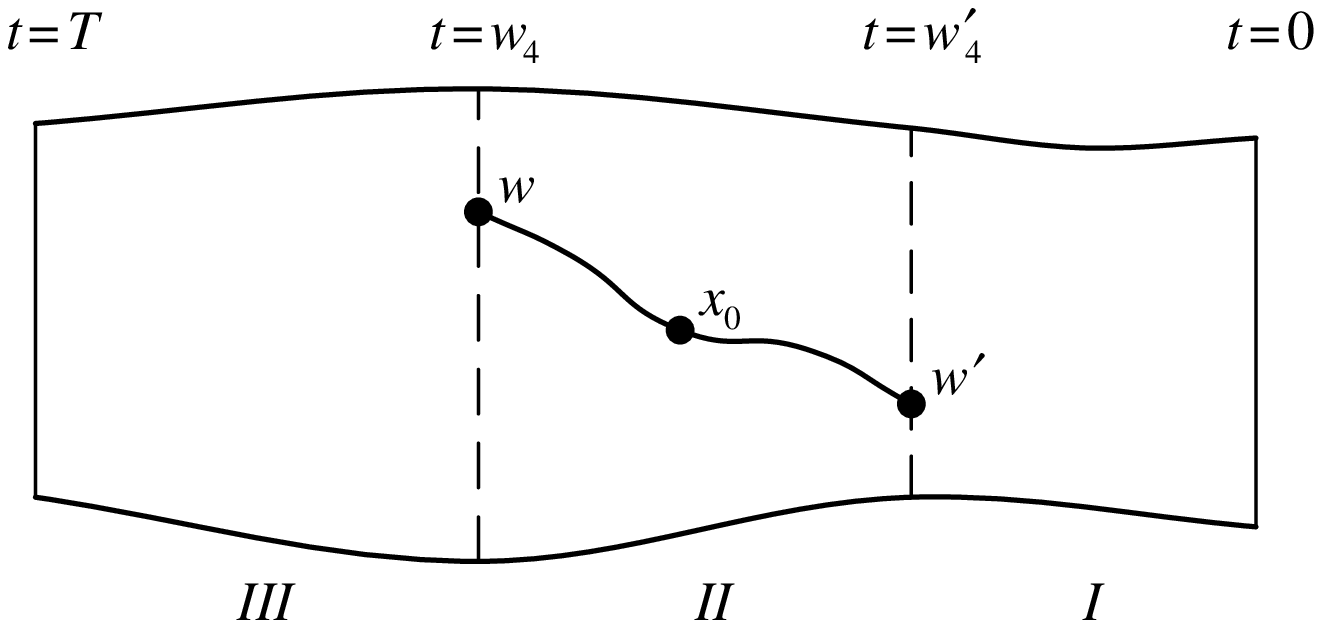}}
\vskip .2cm
\centerline {{\bf Fig. 1} -{\sl The surface {\rm C}}}
\vskip .2cm
$$G^{(S)}_C({\bf r}(t_1),{\bf r}(t_2);t_1-t_2)=
\int {\cal D}{\bf r}(t)\,\exp \left[ {{-m}\over {2}}\int^{t_1}_{t_2}{\rm
d}t\,\dot {\bf r}^2
+C_F\alpha_s\,\int^{t_1}_{t_2}{{{\rm d}t}\over {r(t)}}\right] ,$$

for quarks of equal mass (so that the reduced mass is $m/2$). In region ({\it
II}), however, we
find the octet Green's function,
$$G^{(8)}_C({\bf r}(w_4),{\bf r}(w'_4);w_4-w'_4)=
\int {\cal D}{\bf r}(t)\,\exp \left[ {{-m}\over {2}}\int^{w_4}_{w'_4}{\rm
d}t\,\dot {\bf r}^2
-{{\alpha_s}\over {2N_C}}\,\int^{w_4}_{w'_4}{{{\rm d}t}\over {r(t)}}\right] .$$
\noindent Inserting now this into (10) yields the correction to the total
Green's function,
$G=G^{(0)}+\delta G$:
$$\delta G={{-g^2}\over {2}}\int {\rm d}^3r(w_4)G^{(S)}(r(T),r(w_4);T-w_4)
\int {\rm d}^3r(w'_4)G^{(8)}(r(w_4),r(w'_4);w_4-w'_4)$$
$$\times \int {\rm d}\sigma_{\mu \nu}(w) \int {\rm d}\sigma_{\rho \lambda}(w')
\langle G_{\mu \nu}(w)G_{\rho \lambda}(w')\rangle G^{(S)}(r(w'_4),r(0);w'_4).
\eqno (12)$$
At this point it is convenient to specify the surface inside contour $C$,
 which we do by connecting the points $z(t)$ and $\bar {z} (t)$
 by a straight line. In the {\it c.m.} system,
$${\rm d}\sigma_{\mu \nu}=a_{\mu \nu} {\rm d}z_0 {\rm d}\beta,$$
and, in the nonrelativistic approximation, the $a_{ij}$ may be neglected.
Thus, and as expected, only the chromoelectric piece $\pmbf {$\cal E$}$ of
$G_{\mu \nu}$
 survives in the correlator in Eq. (12). We then expand this correlator in
invariants:
\vskip .1cm
$\langle g^2 {\cal E}_i(x){\cal E}_j(y) \rangle={{1}\over {12}} [\delta_{ij}
\Delta (x-y)+
h_ih_j\,{{\partial D_1}/ {\partial h^2}} ] ,$
\vskip 0cm
\rightline {(13)}
\vskip 0cm
$h=x-y,$
\vskip 0cm
and
$$\Delta (z)=D(z)+D_1(z)+z^2 {{\partial D_1}\over {\partial z^2}}. \eqno (14)
$$
\noindent The invariants $D,\,D_1$ are normalized so that
$$D(0)+D_1(0)=2\pi \langle \alpha_s\,G^2\rangle. \eqno (15) $$
\noindent Inserting then (13) into (12) gives,
$$ \delta G
={{-1}\over {24}}\int {\rm d}^3 r \int {\rm d}^3 r'\int r_i{\rm d}\beta
 \int r'_i{\rm d}{\beta}'\,G_C^{(S)} (r(T),r)\, G_C^{(8)} (r,r')\, G_C^{(S)}
 (r',r(0)).
$$ We next consider the matrix element of $\delta G $ between Coulombic states,
$|nl \rangle$.
 The resulting expression simplifies if we use the spectral decomposition for
$G_C$:
$$G_C^{(S,8)}(r,r';t)=\langle r|\exp [-H_C^{(S,8)}t]|r'\rangle $$
$$=\sum_k \Psi_k^{(S,8)}(r)\,{\rm e}^{-E_k^{(S,8)} t} \Psi_k^{(S,8)}(r')^*.$$
\noindent Identifying the energy shifts from the relation
$$G=G^{(S)}+\delta G \simeq G^{(S)} (1-T\,\delta E_{nl})$$
valid for $T \rightarrow \infty $ we get
$$\delta E_{nl}={{1}\over {36}}\int {{{\rm d}^3p\,{\rm d}p_4}\over {(2\pi )^4}}
\int {\rm d}\beta {\rm d}{\beta}'\,\widetilde {\Delta}(p)
\sum_k \langle nl|r_i \,{\rm e}^{{\rm i}{\bf p}(\beta -{{1}\over {2}}){\bf
r}}|k(8)\rangle$$
$$\times {{1}\over {E_k^{(8)}-E_n^{(S)}-{\rm i}p_4}}
\langle k(8)|r'_i\,{\rm e}^{{\rm i}{\bf p}(\beta -{{1}\over {2}}){\bf
r}'}|nl\rangle
.\eqno (16)$$
\noindent The states $|k(8)\rangle $ are eigenstates of the octet Hamiltonian,
$$H^{(8)}={\bf p}^2/m+\alpha_s \,/2N_C\,r$$
with eigevalues $E_k^{(8)}$. $\widetilde {\Delta}(p)$
is the Fourier transform of $\Delta (x)$.

Eq. (16) is our basic equation. The correlator $\Delta (x)$ depends on $x$ as
$$\Delta(x)=f(|x|/T_g)$$
 and is expected to decrease exponentially for large $|x|$, in Euclidean space.
The
 correlation length $T_g$ may be related to the string tension,
$$T_g^{-1}\equiv \mu_T=
{{\pi}\over {3\sqrt 2}}\,{{\langle \alpha_s\,G^2\rangle^{1/2}}\over
{\sigma^{1/2}}}
\simeq 0.35\,{\rm GeV}. \eqno (17) $$
(for the derivation of this equation, see below). We have now two regions.
 For very heavy quarks, and small $n$,
 $$\mu_T \ll |E_n^{(S)}|=m(C_F\widetilde {\alpha}_s)^2/4n^2.$$
\noindent Then we approximate $\Delta (x)\sim\, { constant}$, and hence
 $\widetilde {\Delta}(p)\sim \delta_4(p)$ so that
$$\delta E_{nl}={{\pi \langle \alpha_s\,G^2\rangle}\over {18}}
\sum_k{{\langle nl|r_i|k(8)\rangle\,\langle k(8)|r_i|nl\rangle}
\over {E_k^{(8)}-E_n^{(S)}}}$$
$$={{\pi \langle \alpha_s\,G^2\rangle}\over {18}}
\langle nl|r_i{{1}\over {H^{(8)}-E_n^{(S)}}}r_i|nl\rangle. \eqno (18) $$
\noindent This is the equation obtained in refs. 1,2 and, upon calculating the
{\it r.h.s.} of (18) one indeed finds the nonperturbative piece of Eq. (1).

The improvement over (18) represented by (16) lies in that it involves the
correlator
$\langle {\cal E}_i(x){\cal E}_j(y)\rangle$, {\rm i.e.}, one takes
 account of nonlocality of the correlator. To implement this
 it is convenient to distinguish two regimes: {\it i}) We consider that $\mu_T$
is smaller than $1/a=mC_F\widetilde {\alpha}_s/2$, but one has $\mu_T\simeq
|E_n|$. {\it ii})
$\mu_T \gg |E_n|$.

We will first consider case ({\it i}). Then one can neglect
$|{\bf p}|$ in the exponents of Eq. (16) but $\mu_T$ should be kept in the
denominator there.
 Using now an exponential form for $\Delta (x)$ we obtain,
$$\widetilde {\Delta}(p)=
{{3(2\pi )^3\mu_T}\over {(p^2+\mu^2_T)^{5/2}}}\langle \alpha_s\,G^2 \rangle,
\eqno (19) $$
 \noindent  Substituting this into (16) we get the energy shifts,
$$\delta E_{nl}={{\pi \langle \alpha_s\,G^2\rangle}\over {18}}
\langle nl|r_i{{1}\over {H^{(8)}-E_n^{(S)}+\mu_T}}r_i|nl\rangle . \eqno (20) $$
\noindent This is precisely the approximation postulated in refs. 9,10.
Clearly, as $T_g \rightarrow \infty$ ($\mu_T \rightarrow 0$), (20) reproduces
 (18); but, as we will see, (20) represents an important improvement
over (18) both from the conceptual and the phenomenological point of view in
the intermediate distance region.

Then we turn to the regime ({\it ii}), $\mu_T \gg |E^{(S)}_n|$. In this case
the
 velocity tends to zero, the nonlocality of the interaction tends to
zero as compared to the quark rotation period
 (which in the Coulombic approximation would be $T_q=1/|E^{(S)}_n|$),
 and the interaction may therefore be
described by a local potential. In fact: considering Eq. (16), it now turns out
 that we can neglect both $E^{(S)}_n$ and the kinetic energy term in
$E^{(8)}_k$
(indeed, all of it) as compared to $-{\it i}p_4$. Then one gets
$$\delta E_{nl} \simeq {{1}\over {36}}\langle nl|
\int_0^{r_i(w_4)}{\rm d}w_i \int_0^{r_i(w'_4)}{\rm d}w'_i \int_0^{\infty}{\rm
d}(w_4-w'_4)
\Delta(w-w')|nl\rangle $$
and in the limit in which we are now working we can approximate
$r_i(w'_4)\simeq r_i(w_4)$.
 One obtains here the matrix elements of a local potential which may be written
in terms of
$D,\,D_1$:$^{[7]}$
$$U(r)=
{{1}\over {36}}\Bigl\{2r\int_0^r {\rm d}\lambda \int_0^{\infty} {\rm d}\nu
D(\lambda ,\nu )$$
$$+\int_0^r \lambda {\rm d}\lambda \int_0^{\infty} {\rm d}\nu [-2 D(\lambda
,\nu )+
 D_1(\lambda ,\nu )] \Bigr\} .$$
\noindent Note that both $D, \,D_1$ depend on $\lambda ,\nu$ through the
combination
$\lambda^2 +\nu^2$.  At large $r$, and as this equation shows, $U(r)$ behaves
like
$$U(r) \simeq \sigma r-{\rm Constant},\,
\sigma={{1}\over {72}}\int {\rm d}x_1 {\rm d}x_2\,D(x_1,x_2).$$
\noindent If we now used the ansatz (19), we would obtain the announced
relation
between $\mu_T$ and $\sigma$. Of course in this situation the strategy of
treating
 the effects of the gluon condensate as a perturbation of the Coulombic
potential is no more
appropriate. One should rather take $U(r)$, together with the Coulombic
potential,
as part of the unperturbed Schr\"odinger equation. We leave the subject here
referring to the various existing analyses$^{[5,11,12]}$ for details.(In
particular, refs. 11
 are the ones closer in spirit to the work here in what regards the treatment
of the nonperturbative
effects, while ref. 12 incorporates radiative corrections to a phenomenological
long distance potential)
\vskip 1.5cm
{\bf Phenomenology}
\vskip .2cm
For the phenomenological analysis we will generalize Eq. (1) by writing,
$$M(n,l)=2m \Bigl\{ 1-{{C_F\widetilde {\alpha}_s^2}\over {8n^2}}$$
$$-\left[ \log{{n\mu}\over {mC_F\widetilde {\alpha}_s}}+\psi (n+l+1)\right]
\,{{C_F^2\beta_0 \widetilde {\alpha}_s \alpha_s^2}\over {8\pi n^2}}
+{{\pi \epsilon_{nl}(\mu_T) n^6\,\langle \alpha_s\,G^2 \rangle }
\over {2(mC_F\widetilde {\alpha}_s)^4}}\Bigr\}, \eqno (21) $$
where the $ \epsilon_{nl}(\mu_T) $ are obtained solving Eq. (20); for $\mu_T
\rightarrow 0$,
 $ \epsilon_{nl}(0)= \epsilon_{nl} $, this last being the quantities given in
Eqs. (1), (2).

Unlike in the case $\mu_T=0$, where a closed expression could be found for the
$\epsilon_{nl} $,
 the $\epsilon_{nl}(\mu_T) $ may only be computed numerically. However,
 a fairly precise evaluation may be obtained
by neglecting the potential $\alpha_s\,/2N_Cr$ in (20), then working with
$p$-space Coulomb
functions. To a very tolerable $\sim $5\% accuracy it follows that we may
approximate,
$$ \epsilon_{nl}(\mu_T) \simeq {{ \epsilon_{nl}}\over {1+\rho_{nl} \eta_n}},
\,\,\eta_n\equiv \left ({{2n}\over {C_F\tilde {\alpha}_s}} \right
)^2{{\mu_T}\over {m}},
 \eqno (22) $$
where,
$$\rho_{10}=0.62,\,\rho_{20}=0.76,\,\rho_{30}\approx 0.9,\,\rho_{21}=0.70.$$
\noindent Substituting into Eq. (21) then gives us a very explicit
generalization of Eq. (1), valid
in the intermediate region $\mu_T \sim |E_n^{(S)}|$.

For the numerical calculation we proceed as follows. For $b\bar {b}$ we take
the optimum values for the
renormalization point $\mu$ given in ref. 4 (which were obtained neglecting
$\mu_T$). We therefore choose,
$$\mu=1.5\,{\rm GeV}, {\rm for}\,n=1;\,\mu=0.95\,{\rm GeV},\,{\rm for}\,n=2.$$
\noindent For mixed $n$ we take the value corresponding to the {\it smaller}
$n$.
 For $c \bar {c}$ we, somewhat arbitrarily, choose $\mu=0.95\,{\rm GeV}$. For
the
basic QCD parameters we take
$$\Lambda(n_f=4, 2\,{\rm loops})=200\,{\rm MeV},\,\langle \alpha_s\,G^2 \rangle
=0.042\, {\rm GeV}, $$
\noindent and thus the corresponding values of $\alpha_s$ are
$\alpha_s (1.5\, {\rm GeV})=0.27,\,\alpha_s (0.95\, {\rm GeV})=0.35. $

 We will not consider varying these quantities. A variation of $\Lambda$ can
be largely compensated by a corresponding variation of $\mu$ and
likewise, and because the $\epsilon_{nl}(\mu_T)$ depend almost
 exactly on the ratio $\langle \alpha_s\,G^2 \rangle /\mu_T,$ a
variation of the condensate may be balanced by a compensating variation
 of the correlation time, $T_g=\mu_T^{-1}$.

We now have two possibilities: fit $\mu_T$ to each individual
 splitting, and compare the resuls
 among themselves and with the one coming from the string tension; or take
$\mu_T$ from the last, Eq. (17)
 and then {\it predict} the splittings. If we do the first we find
$$\mu_T=0.40\,{\rm GeV}\,({\rm 2S-1S}),\,\mu_T=0.76\,{\rm GeV}\,({\rm
3S-2S}),\,
\mu_T=0.59\,{\rm GeV}\,({\rm 2S-2P}). \eqno (23) $$
For the $c\bar {c}$ case we only consider the 2S-1S splitting, where we get
$$\mu_T=1.23\,{\rm GeV}. \eqno (24) $$
Clearly, the more reliable calculation is that of the  2S-1S $b\bar {b}$
splitting: not only the
 radiative corrections are known (unlike for the 2S-2P case) but also it falls
inside the
 conditions of regime ({\it i}), unlike the 3S-2S splitting and, even more,
 the  2S-1S  $c\bar {c}$ one. It is then gratifying that the value of
$\mu_T$ that follows from the $b\bar {b}$  2S-1S splitting, $\mu_T=0.4$ GeV,
 is the one which is in better agreement with the value $\mu_T=0.32$ GeV
obtained with the comparison with
 the linear potential, Eq. (17).

If we now choose the second possibility, we fix $\mu_T$.
 The corresponding results are summarized in Table I.
\vskip .1cm
$$\matrix { {\sl splitting} &\mu_T=0 & \mu_T=0.32 & \mu_T=0.40 & {\sl exp.} \cr
& & & & \cr
{\rm 2S-1S}\,(b\bar {b})&479^a &590 & 522 & 558\,{\rm MeV} \cr
{\rm 2S-2P}\,(b\bar {b})&181^a &162 & 147 & 123\,{\rm MeV} \cr
{\rm 3S-2S}\,(b\bar {b})&4\,570 &748 & 614 & 332\,{\rm MeV} \cr
& & &  \cr
{\rm 2S-1S}\,(c\bar {c})&9\,733 &1\,930 & 1\,626 & 670\,{\rm MeV} } $$
\vskip .1cm
\centerline {{\bf Table I}.- Predicted splittings, and experiment.}
\vskip 0cm
\centerline { ($a$): Values from ref 4, with $\mu \simeq 0.95$ for both}
\vskip .2cm

The $n=2$ $b\bar {b}$ states are certainly better described than
 with the approximation $T_G=\infty$ of refs. 1, 2, 4. Particularly important
is the
fact that inclusion of the finite correlation time {\it stabilizes} the
calculation. For
example, if we had taken $\mu =1.5$, {\it and} $\mu_T=0$,
 for the 2S-1S splitting for bottomium,
 we would have obtained the absurd value of 1 944 {\it GeV}.
 In what respects the $n=3$ and the $n=2\,c\bar {c}$ states
the improvement is marginal, in the sense that the basic assumption,{\it viz.},
that one can treat the
nonperturbative effects at leading order fails, as is obvious from the figures
in
 the column "$\mu_T=0$" in Table I. Indeed these states
fall clearly in regime ({\it ii}) and should therefore
 be better described with a local potential as discussed extensively
 in the existing literature, of which we, and for illustrative purposes, single
out ref. 11, where the
nonperturbative effects (including spin-dependent splittings) are treated with
methods
 like ours, but where radiative corrections are ignored; or ref. 12 where
radiative
corrections are incorporated but the confining potential is introduced
phenomenologically.

We would like to end this note with a few words on extensions of this work. An
obvious one
is to include the treatment of spin effects, and a calculation of the wave
functions.
 Then it would be very desirable to evaluate the radiative corrections to the
 nonperturbative terms, as this would greatly diminish the dependence of these
terms on the renormalization point, $\mu$, thereby substantially increasing
 the stability of the calculation. Finally, and to be able to extend the
calculation to
intermediate distances with success, one should abandon the treatment of
nonperturbative effects at
first order: an iterative approach should certainly yield better results.

\vskip 3cm
{\bf Acknowledgements}.-
\vskip .2cm

We are grateful to CICYT, Spain, for partial financial support. One of us (Yu.
A. S.) would like to
 acknowledge the hospitality of the Universidad Aut\'onoma de
 Madrid, where most of this work was done.
\vfill
\eject
{\bf REFERENCES}
\vskip .25cm
1.-H. Leutwyler, Phys. Lett., {\bf 98B}, 447 (1981)

2.-M. B. Voloshin, Nucl. Phys., {\bf B154}, 365 (1979); Sov. J. Nucl. Phys.,
{\bf 36}, 143 (1982)

3.-F. J. Yndur\'ain, {\sl The Theory of Quark and Gluon Interactions},
Springer, 1993

4.-S. Titard and F. J. Yndur\'ain, Phys. Rev. {\bf D49}, 6007 (1994); FTUAM
94-6, in press in Phys. Rev.
FTUAM 94-34.

5.-E. Eichten {\it et al.}, Phys. Rev., {\bf D21}, 203 (1980); W. Lucha, F.
Sch\"oberl and D. Gromes,
Phys. Rep., {\bf C200}, 128 (1991)

6.-I. I. Balitsky, Nucl. Phys., {\bf B254}, 166 (1985)

7.-H. G. Dosch, Phys. Lett. {\bf B190}, 177 (1987); Yu. A. Simonov,
 Nucl. Phys., {\bf B307}, 512 (1988) and {\bf B324}, 56 (1989);
H. G. Dosch and Yu. A. Simonov, Phys. Lett., {\bf B205}, 339 (1988); H. G.
Dosch and M. Schiestl,
Phys. Lett., {\bf B209}, 85 (1998). For a review, see Yu. A. Simonov, Sov. J.
Nucl. Phys., {\bf 54},
 192 (1991)

8.-J. A. Tjon and Yu. A. Simonov, Ann. Phys (N.Y.), {\bf 228}, 1 (1993)

9.-A. Kr\"amer, H. G. Dosch and R. A. Bertlmann, Phys. Lett., {\bf B223},
 105 (1989); Fort. der Phys., {\bf 40}, 93 (1992)

10.-M. Campostrini, A. Di Giacomo and S. Olejnik, Z. Phys., {\bf C31}, 577
(1986)

11.-A. M. Badalian and V. P. Yurov, Yad. Fiz., {\bf 51}, 1368 (1990);
Phys.Rev.,
{\bf D42}, 3138 (1990)

12.-See, {\it e.g.} S. N. Gupta, S. F. Radford and W. W. Repko,
 Phys Rev., {\bf D26}, 3305 (1982)

\vskip 0cm

\end